\documentstyle[draft,aps,epsfig]{revtex}
\def\rxn{$e^+e^-\rightarrow hadrons$}
\def\ftrho{F^{(3)}_\rho}
\def\ferho{F^{(8)}_\rho}
\def\ftomega{F^{(3)}_\omega}

\def\ftphi{F^{(3)}_\phi}

\def\ms{$m_s$}
\begin{document}
\draft
\title{Isospin Breaking and the Extraction of
$m_s$ from the $\tau$-Decay-Like \\ Vector Current Sum Rule}
\author{Kim Maltman\thanks{e-mail: maltman@fewbody.phys.yorku.ca}}
\address{Department of Mathematics and Statistics, York University, \\
          4700 Keele St., North York, Ontario, CANADA M3J 1P3 \\ and}
\address{Special Research Center for the Subatomic Structure of Matter, \\
          University of Adelaide, Australia 5005}
\maketitle
\begin{abstract}
Narison's $\tau$-decay-like sum rule for determining the strange
quark mass 
is re-investigated, taking into account 
isospin-breaking corrections in the extraction of the input spectral functions
from {\rxn} data.
The corrections,
estimated using experimental data on vector meson electromagnetic
decay constants and a QCD sum rule analysis of the
$38$ vector current correlator, are shown to be especially large
for the isoscalar case.
The reason such large corrections are natural
is also explained.
Due to the high degree of cancellation in the original sum rule,
the effect of these corrections on the determination of $m_s$ is significant.
A new central value
$m_s=113-138\ {\rm MeV}$ is found, 
in the $\overline{MS}$ scheme at $1$ GeV$^2$,
with significant (asymmetric) errors associated with
errors in the input experimental data.
\end{abstract}

\section{Introduction}
The usually quoted values for the light 
($u,d,s$) current quark masses
are obtained by a sum rule analysis of the correlator of either
(for $m_u+m_d$) the product of two divergences
of the isovector axial vector current\cite{bpr,prades97} or (for 
$m_s$) the product of two divergences of the
strangeness-changing vector current\cite{jm,cps}.  One of the
potential problems with this approach is that, in both
cases, continuum contributions to the hadronic side of the sum rule
are large, but the continuum part of the hadronic spectral function
is not known experimentally.
The extracted
quark masses thus depend crucially for their reliability 
on that of the theoretical ansatze
for the unmeasured continuum spectral functions, which
are constructed by analogy
with an extreme form of the vector meson dominance 
(VMD) treatment of the
vector isovector channel wherein
the continuum spectral function is
modelled as a sum of Breit-Wigner terms
whose overall normalization is adjusted to produce the desired value at
continuum threshold.
The threshold value is estimated, for the pseudoscalar channel, 
using tree-level chiral perturbation theory 
(ChPT)\cite{bpr},
and for the $S=-1$ scalar channel, 
by extrapolating $K_{\ell 3}$ data
using the Omnes representation, together with experimental
data on the $K\pi$ scattering 
phases\cite{jm}.  Since the relevant thresholds are many
resonance widths away from the poles, 
the assumed $q^2$-dependence of the resonance widths
is also a crucial input.
Taking the $m_s$ analysis 
to be specific, the ``standard'' form of the $q^2$-dependent $s$-wave width
is employed.  This form results from assuming the scalar couplings of
the resonances in question to $K\pi$ are $q^2$-independent, a somewhat
dangerous assumption in the scalar channel.

The above assumptions have generally been considered plausible
because an analogous version is known to allow a successful description
of $e^+e^-\rightarrow\pi^+\pi^-$ cross-sections. The analogy is, however,
potentially dangerous\cite{bgm,rhostuff,colangelo}.
In Ref.~\cite{colangelo}, for example,
the $K\pi$ portion of the scalar, $S=-1$ spectral function is
obtained {\it for all $s$} using the Omnes representation of the
timelike scalar $K\pi$ form factor, with $K_{e3}$ and $K\pi$ phase
data as input.  The resulting spectral function rises much faster
just above threshold, and reaches a much lower (by a factor of $\sim 3$) 
$K_0^*(1430)$
peak height, than does the model version, obtained using an assumed
resonant spectral shape normalized at threshold, employed in
Refs.~\cite{jm,cps}.
Above $s\sim 2\ {\rm GeV}^2$, where $K\pi$ states no longer
dominate the spectral function, a full determination of
the relevant spectral function via the method of Ref.~\cite{colangelo}
is no longer possible;
below $s\sim 2\ {\rm GeV}^2$, however, this result clearly
demonstrates the existence of
potentially
large uncertainties in the
``resonance-saturation/threshold-normalization'' ansatze for the unmeasured
continuum spectral functions.  This
suggests it is preferable to employ a sum
rule involving experimentally-determinable spectral data.  
The sum rule for $m_s$ proposed by
Narison\cite{narison95} is of this type.

Narison's idea is to consider the difference of
isovector and hypercharge vector current correlators,
$\Pi^{33}-\Pi^{88}$, where $3$ and $8$ are $SU(3)_F$
labels, and
$\Pi^{aa}$ is
defined by
\begin{equation}
i\int \ d^4x\ e^{iq\cdot x}
<0\vert T\left( J^a_\mu (x) J^a_\nu (0)\right)\vert 0>
= (q_\mu q_\nu -q^2 g_{\mu\nu})\Pi^{aa}(q^2).
\end{equation}
In the operator product expansion (OPE) for this difference, each
term necessarily involves the flavor-breaking parameter $m_s-\hat{m}$
(where $\hat{m}=(m_u+m_d)/2$).  
Narison proposes integrating the corresponding 
spectral function, 
weighted as for $\tau$ decay kinematics, from
threshold $q^2=4m_\pi^2$ up to a mock $\tau$ mass, $q^2=m_t^2$.
As for inclusive
hadronic $\tau$ decays\cite{tau1,tau2,bnp,ldp,np93,tau6},
analyticity properties allow one to re-write this integral
as a correspondingly weighted integral
of the correlator difference over a circular contour of
radius $m_t^2$ in the complex $q^2$ plane.  
For large enough $m_t$, this alternate representation simultaneously
suppresses both contributions from the region of the contour near
the positive real axis where perturbative QCD (pQCD) becomes
unreliable, and non-perturbative
contributions in the OPE relative to perturbative 
ones\cite{tau1,tau2,bnp,ldp,np93,tau6}.  
In addition, if one
ignores isospin breaking, the relevant spectral functions are 
experimentally determinable in $e^+e^-\rightarrow hadrons$. 
Narison's original treatment\cite{narison95}, which
neglected isospin breaking, produced 
$m_s$ values compatible with those of the conventional
analyses\cite{jm,cps} ($197\pm 29\ {\rm MeV}$ 
in the $\overline{MS}$
scheme, at a scale of $1$ GeV$^2$, c.f.
$205\pm 19$ MeV\cite{cps}).
Since, however, there is a high degree of 
cancellation between the 33 and 88 contributions to
the sum rule, it is important to consider the possibility
of isospin-breaking corrections to the hadronic spectral functions.
We investigate this question in the present paper.

\section{Narison's $\tau$-Decay-Like Sum Rule for \ms }

Narison's method is based on an analogy with analyses of 
the inclusive $\tau$ decay ratio
\begin{equation}
R_\tau^{I=1}=\frac{\Gamma_{I=1}\left( \tau\rightarrow
\nu_\tau hadrons (\gamma )\right)}{\Gamma \left(\tau\rightarrow
e\nu_\tau\bar{\nu}_e (\gamma )\right)},
\end{equation}
(for concreteness, we consider
$\tau$ decays mediated by the charged weak isovector vector
current).  $R_\tau^{I=1}$ is given by an integral over
the $J=0,1$ scalar spectral functions of the vector
isovector correlator, weighted by the appropriate kinematic 
factors\cite{tau1,tau2,bnp,ldp,np93,tau6}, 
which integral can be converted into one involving
the scalar correlators themselves,
with the same kinematic weights, over the counterclockwise
oriented circular contour of radius $m_\tau^2$ in the complex
$s=q^2$ plane.  Narison considers the difference
$R_t^{33} - R_t^{88}$, where $R_t^{aa}$ ($a=3,8$) results from
replacing $m_\tau$ by
a variable mass, $m_t$, and the isovector
current in $R_\tau^{I=1}$ by $J^a_\mu$.
This difference can be expressed in either the ``hadronic'' or
``contour integral''
representations
\begin{eqnarray}
\left[ R_t^{aa}\right]_{had} &=& 12\pi^2\, \vert V_{ud}\vert^2 S_{EW}\, 
\int_0^{m_t^2}\, {\frac{ds}{m_t^2}}\,
\left( 1-{\frac{s}{m_t^2}}\right)^2
\left( 1+{\frac{2s}{m_t^2}}\right)\, \rho^{aa}(s)
\label{hadronaa} \\
\left[ R_t^{aa}\right]_{contour} &=& 6\pi i\, \vert V_{ud}\vert^2 S_{EW}\, 
 \int_{\vert s\vert =m_t^2}\, {\frac{ds}{m_t^2}}\,
\left( 1-{\frac{s}{m_t^2}}\right)^2
\left( 1+{\frac{2s}{m_t^2}}\right)\, \Pi^{aa}(s)\ ,
\label{OPEaa}\end{eqnarray}
where $V_{ud}$ is the $ud$ CKM matrix element,
$S_{EW}=1.0194$ is the sum of the leading-log electroweak $\tau$ decay
corrections\cite{ms88},
$\rho^{33}(s)$ and
$\rho^{88}(s)$ are the isovector and isoscalar spectral functions and,
for sufficiently large $m_t$, $\left[ R_t^{aa}\right]_{contour}$
can be evaluated using the
OPE. 
The sum rule for $m_s$ results from
equating Eqs.~(\ref{hadronaa}) and (\ref{OPEaa})\cite{narison95}.

For the OPE side one has, with $D$ labelling operator 
dimension\cite{narison95},
\begin{equation}
\left[ R^{33}_t-R^{88}_t\right]_{OPE}=
\vert V_{ud}\vert^2 S_{EW}\sum_{D=2,4,\cdots}\left(\delta^{(D)}_{uu}
-\delta^{(D)}_{ss}\right)\ .
\label{OPEside}\end{equation}
The $D=2$ terms result from the mass-dependent perturbative
contributions to the correlators which, for a flavor-diagonal vector
current of flavor $i$, are\cite{chetyrkin}:
\begin{eqnarray}
\Delta^{(mass)}_{ii}(s)&=& -{\frac{3}{2\pi}}{\frac{
\bar{m}_i(Q^2)}{Q^2}}\left[ 1+{\frac{8}{3}} a(Q^2)
+\left( {\frac{17981}{432}} +{\frac{62}{27}}\zeta (3)
-{\frac{1045}{54}}\zeta (5)\right) a(Q^2)^2\right] \nonumber \\
&&+{\frac{1}{12\pi^2}}a(Q^2)^2 \left( 32-24\zeta (3)\right)
\sum_k {\frac{\bar{m}_k(Q^2)}{Q^2}}\ +\ {\cal O}(a^3)
\label{corrmass}
\end{eqnarray}
with $a(Q^2) =\alpha_s(Q^2)/\pi$
and $\bar{m}_j(Q^2)$ the running coupling and running mass of quark $j$, both
at scale $\mu^2=Q^2=-s$, in the $\overline{MS}$ scheme.
Expanding $a(Q^2)$ in terms of $a(m_t^2)\equiv a$
and $\bar{m}_j(Q^2)$ in terms of $a$ and
$\bar{m}_j(m_t^2)\equiv \bar{m}_j$,
and performing the resulting
elementary logarithmic integrals, one obtains
\begin{equation}
\delta^{(2)}_{uu}-\delta^{(2)}_{ss}=
{\frac{12\left( \bar{m}^2_s
-\bar{m}^2_u\right)}{m_t^2}}\left( 1+f_1 a+f_2 a^2 +f_3 a^3+\cdots 
\right)\label{D2}
\end{equation} 
where $f_1 =13/3$, $f_2 =30.5846$, and $f_n$,
$n>2$ are unknown.
Here $f_1$ differs from that of Narison
($f_1=11/3$), but is in agreement with the equal mass case of
Eq.~(3.8) of Ref.~\cite{bnp}.  Narison does not quote a value
for $f_2$.
Similarly, for the $D=4,6$ contributions, one finds\cite{bnp,narison95},
with $\rho$ representing the deviation of the light 4-quark
condensate from its vacuum saturation value,
\begin{eqnarray}
\delta^{(4)}_{uu}-\delta^{(4)}_{ss}&=&{\frac{36\pi^2a^2}{m_t^4}}
\left[ < m_s\bar{s} s>-< m_u \bar{u} u>\right]
+{\frac{36}{m_t^4}}\left( \bar{m}_u^4-\bar{m}_s^4\right)\label{D4} \\
\delta^{(6)}_{uu}-\delta^{(6)}_{ss}&=&
{\frac{1792\pi^3}{27}}
{\frac{\rho\alpha_s(m_t^2)}{m_t^6}}\left[ <\bar{u}u>^2-<\bar{s}s>^2\right]\ .
\label{D6}\end{eqnarray}
Note that the perturbative series in 
Eq.~(\ref{D2}) converges much more slowly than does the analogous
mass-independent perturbative contribution to inclusive $\tau$
decay, $P_{incl}(a)
=1+a+5.2023 a^2 + 26.366 a^3+\cdots$.
For $m_t=m_\tau$, e.g.,
using $\alpha_s(m_\tau^2 )=0.351\pm 0.016$\cite{ALEPHalphas},
$P_{incl}\left( a(m_\tau^2 )\right)=1+0.1114+0.0645+0.0365+\cdots$, while
$P_{mass}\left( a(m_\tau^2 )\right)=1+0.4828+0.3981+\cdots $.
Assumptions about the convergence of $P_{mass}(a)$
based by analogy on the
behavior of $P_{incl}(a)$ can, thus, not be expected to be reliable.

For the hadronic side of the sum rule, 
neglecting isospin breaking, one has, using the narrow width
approximation for the isoscalar contributions\cite{narison95},
\begin{eqnarray}
R^{33}_t &=& {\frac{3\vert V_{ud}\vert^2 S_{EW}}{2\pi\alpha_{EM}^2}}\, 
\int_0^{m_t^2}\, ds\, 
\left( 1-{\frac{s}{m_t^2}}\right)^2\left( 1+{\frac{2s}{m_t^2}}\right)
{\frac{s}{m_t^2}}\sigma^{(I=1)}_{e^+e^-\rightarrow hadrons}\label{33had} \\
R^{88}_t&=&{\frac{18\pi\vert V_{ud}\vert^2 S_{EW}}{\alpha_{EM}^2}}\, \left[
\left( 1-{\frac{m_\omega^2}{m_t^2}}\right)^2
\left( 1+{\frac{2m_\omega^2}{m_t^2}}\right){\frac{m_\omega 
\Gamma_{\omega\rightarrow e^+e^-}}{m_t^2}}+(\omega\rightarrow\phi )
+\cdots\right]\label{88had}
\end{eqnarray}
where $+\cdots$ refers
to continuum and higher resonance
contributions, which are small for $m_t$ less than $\sim 1.6$ GeV,
and have been estimated by Narison\cite{narison95}.  
Unfortunately, there turns out, numerically, to be a
high degree of cancellation (to the $10-15\%$ level) in
$R^{33}_t - R^{88}_t$.  With $m_t=1.4,\ 1.6$ GeV, and
Narison's evaluation of the
hadronic side, for example,
\begin{eqnarray}
\left[ R^{33}_t-R^{88}_t\right]_{m_t=1.4\ {\rm GeV}}&=&
(1.853\pm 0.072)-(1.581\pm 0.066)=0.272\pm 0.098 \\
\left[ R^{33}_t-R^{88}_t\right]_{m_t=1.6\ {\rm GeV}}&=&
(1.793\pm 0.070)-(1.626\pm 0.069)=0.167\pm 0.098\ .
\end{eqnarray}
Isospin breaking at the few $\%$ level in the individual
terms is, therefore,
not necessarily negligible in the difference, especially
if, as one would expect, e.g., for $\rho$-$\omega$ mixing effects,
the signs of the effect were to be opposite in the two cases.

We now describe the input to the present version of analysis of
the Narison sum rule.
Modifications to the OPE side of the sum rule are minor.  First,
we use Narison's updated
value of the $D=6$ 4-quark condensate,
$\rho\alpha_s <\bar{u}u>^2=(5.8\pm 0.9)\times 10^{-4}\ 
{\rm GeV}^6$\cite{narison952},
and, in evaluating the difference of light and strange quark
terms, allow $<\bar{s}s>/<\bar{u}u>$
to vary between $0.7$ and $1$, as in Ref.~\cite{colangelo}.  
Second, for the (very small) $D=4$
terms, we take the light quark condensate
to be given by the GMOR relation, and the strange quark 
condensate by the kaon version thereof,
multiplied by a factor, $c_K$, (varied
between $0.5$ and $1$ to account
for possible $SU(3)_F$ breaking).
The ratio $m_s/\hat{m}=24.4\pm 1.5$, determined
from ChPT to 1-loop\cite{leutwyler96}, is also used.
Third, since 
the 4-loop $\beta (a)$\cite{beta4} and $\gamma (a)$\cite{gamma4} functions
are now available, we employ throughout the 4-loop expressions for 
$a(Q^2)$ and $\bar{m}(Q^2)$,
fixing the 3-flavor scale parameter, $\Lambda_3$ from the recent
ALEPH $\tau$ decay data analysis\cite{ALEPHalphas}.
The slow convergence noted above for
$P_{mass}(a)$ at $m_t=m_\tau$ 
is even more accentuated at lower scales
(e.g., for $m_t=1.2,\ 1.4$ GeV, $P_{mass}=1+0.672+0.736+\cdots$
and $1+0.580+0.548+\cdots$, respectively).
An estimate of the ${\cal O}(a^3)$ term in $P_{mass}(a)$ thus appears
crucial.  We estimate $f_3$ using the
procedure of Ref.~\cite{cks} (CKS).
For the three cases where the ${\cal O}(a^3)$ coefficient
of quadratically-mass-dependent observables are known, the 
resulting estimates are accurate to $\pm 25$\cite{cks}.  
The two possible versions, labelled FAC and PMS,
yield $[f_3]^{FAC}=288.0$
and $[f_3]^{PMS}=290.1$.  In contrast, [1,1] and [0,2]
Pade estimates yield $[f_3]^{[1,1]}=215.9$ and
$[f_3]^{[0,2]}=183.7$.  To be conservative,
we take $f_3=200\pm 200$.

On the hadronic side the
major change 
is that we now make isospin-breaking 
corrections to the extracted isovector and isoscalar 
spectral functions.
These turn out to be
quite large.  Such corrections are unavoidable
in the isoscalar case, where {\rxn} is the only source of
experimental data. In the isovector case,
we will combine {\rxn} and $\tau$ decay data in order to
reduce the errors on $R^{33}_t$.  For more
details on the isospin breaking corrections to the 
{\rxn} data, see Ref.~\cite{krmcvc}.
The necessity of such corrections is obvious.  Just as
isospin breaking is observed through $\rho$-$\omega$ interference
in $e^+e^-\rightarrow hadrons$, so the vector meson electromagnetic (EM)
decay constants, $F^{EM}_V$,
will have both isospin-conserving and isospin-violating
pieces.  Defining $F^a_V$ by
$<0\vert J^{a}_\mu \vert V,\lambda >=F^{a}_V m_V \epsilon_\lambda$
for $a=3,8$, we see that
$F^{EM}_V=F^3_V+{\frac{1}{\sqrt{3}}}F^8_V$.
$F^8_\rho$, $F^3_\omega$ and $F^3_\phi$
vanish in the isospin limit, but will be non-zero
in the real world.  The $\rho$ contribution
to the EM spectral function then contains an isospin-conserving
piece proportional to $(\ftrho )^2$ and an isospin-violating
piece proportional to $\ftrho\ferho$.
The latter (associated with the 38 portion of $\rho^{EM}$)
must be excluded in determining the $\rho$ contribution
to $\rho^{33}$.
Similarly, contributions proportional to $F^8_V F^3_V$ ($V=\phi ,\omega$)
should be removed from the physical
EM widths to obtain 
the $\omega$ and $\phi$
contributions to the 88 spectral function.
The evaluation of the (unmeasured) isospin-breaking 
decay constants is accomplished
by performing a QCD sum rule analysis of the
mixed-isospin correlator
$<0\vert T\left(J^3_\mu J^8_\nu\right)\vert 0>$\cite{krmcvc,ijl}, for which
the resonance contributions are directly proportional to
the product $F^3_V F^8_V$.  Combining the results for these
products with the experimental values for
$F^{EM}_V$, one extracts $F^3_V$ and $F^8_V$
separately.  The sum rule provides good
constraints on the product for the $\rho$ and $\omega$, weak
constraints for the $\phi$, and is insensitive to higher resonance
contributions.  (Details of the analysis, in relation to
CVC tests and the extraction of the sixth order ChPT low-energy
constant, $Q$\cite{gk,krmq}, may be found in Ref.~\cite{krmcvc}.)
One finds
$\ferho = 2.8\pm 1.1$ MeV, $\ftomega = -4.2\pm 1.5$ MeV
and $\ftphi = 0.21 \pm 0.21$ MeV
( c.f., $F^{EM}_\rho = 154\pm 3.6$ MeV, $F^{EM}_\omega = 45.9\pm 0.8$ MeV
and $F^{EM}_\phi = -79.1\pm 2.3$ MeV)\cite{krmcvc}.  These results
satisfy several physical naturalness criteria\cite{krmcvc}.
The $\rho$,
$\omega$ and $\phi$
contributions to the $33$ and $88$ vector current
spectral functions are given by
$\left[ \rho^{aa}(q^2)\right]_{V} = \left[ F^{a}_V\right]^2
\delta_V (s)$
where $\delta_V(s)$, $V=\omega ,\phi$, in the
narrow width approximation, is the usual $\delta$ function, 
while $\delta_\rho (s)$ is the corresponding $\rho$ Breit-Wigner.
The standard extractions, in contrast, are obtained
by replacing $F^{3}_\rho$ with $F^{EM}_\rho$ and $ F^{8}_V$
with $\sqrt{3}F^{EM}_V$ for $V=\omega ,\phi$.  The corrections
necessary to produce the true resonance contributions
to the $33$ and $88$ spectral functions are thus
\begin{eqnarray}
\left[ \frac{F^{(3)}_\rho}{F^{EM}_\rho}\right]^2&=& 0.979\pm 0.0086
\nonumber \\
\left[ \frac{F^{(8)}_\omega}{\sqrt{3}\, F^{EM}_\omega}\right]^2&=& 
1.189\pm 0.065 \nonumber \\
\left[ \frac{F^{(8)}_\phi}{\sqrt{3}\, F^{EM}_\phi}\right]^2&=& 
1.0054\pm 0.0054\ .
\label{corrections}
\end{eqnarray}
The size of the overestimate in the case of the $33$ spectral function
is still noticeably smaller than the $\sim 5\%$ errors on
the {\rxn} cross-sections in the resonance region.  Note that the
scale ($\sim 1\%$) of the isospin-breaking contribution to
$F^{EM}_\rho$ corresponds, as one
might expect (since the assumption of mixing dominance corresponds
to a leading chiral order approximation\cite{krmcvc}), 
to what is obtained by assuming
dominance by $\rho$-$\omega$ mixing, and then evaluating
this mixing following the updated analyses
of $e^+e^-\rightarrow \pi^+\pi^-$ discussed in Refs.~\cite{newrho}.  
Note also that, in this approximation, and
neglecting both the $\rho$ width and the difference of the $\rho$ and $\omega$ 
masses, the $\rho$ and $\omega$ correction terms would cancel
in spectral integrals.
The approximate cancellation of the corrections associated with
the results of Eq.~(\ref{corrections}), when one forms a sum rule involving
the {\it sum} of $\rho$ and $\omega$ contributions, is thus simply
a manifestation of the dominant role of $\rho$-$\omega$ mixing.
Without understanding this point, it is easy to
be misled into thinking that the scale of the individual resonance
isospin-breaking decay constants is set by that of the
isospin-breaking terms in the corresponding OPE representation.
The latter, however, is associated with the sum, which involves
a rather close cancellation, rather than the scale of the individual
terms.
While the $1\%$ correction in the case of the $\rho$ is, as 
just explained, quite natural,
the $\omega$ correction 
to the $88$ spectral function ($\sim 19\%$) might, in contrast,
seem unnaturally large to some readers.
It is, however, rather easy to see that such a large
correction is actually expected in this case.
As is well-known, in the limit that the vector meson nonet
is ideally mixed, but the octet vector current matrix
elements are otherwise given by $SU(3)_F$, the
EM decay constant of the $I=1$ component of the $\rho$, $F_\rho^I$,
is 3 times that of the $I=0$ component of the $\omega$, $F_\omega^I$.
Writing $\rho = \rho_I +\epsilon\, \omega_I$ and 
$\omega =\omega_I -\epsilon\, \rho_I$, 
($I$ denoting the isospin pure states), with
$\epsilon\sim {\cal O}(\delta m)$, the physical EM decay constants
become
\begin{eqnarray}
F^{EM}_\rho&=&F^I_\rho +\epsilon F^I_\omega
\simeq F^I_\rho\left( 1+\frac{\epsilon}{3}\right) \nonumber \\
F^{EM}_\omega&=&F^I_\omega -\epsilon F^I_\rho
\simeq F^I_\omega\left( 1-3\epsilon \right) \, .
\end{eqnarray}
The fractional correction for the $\omega$ should thus
be $\sim 9$ times that for the $\rho$, and of opposite sign.
Both features are present in the results of the sum rule analysis.

In reanalyzing the sum rule, we would, ideally, prefer to work with
$m_t$ near $m_\tau$ where the perturbative series
$P_{mass}(a)$ is better behaved.  Unfortunately, at such
scales, the $\omega^\prime$ contribution to $R^{88}_t$
becomes important.  Since the sum rule
employed to estimate the isospin-breaking
decay constants is insensitive to
the $\rho^\prime$-$\omega^\prime$ region, we are unable
to correct the $\omega^\prime$ contribution
for isospin breaking, and must thus
work at scales below $m_t\sim 1.6$ GeV, where
this contribution is small ($<2\%$\cite{narison95}).
The slow convergence of $P_{mass}(a)$, similarly, forces us to
values of $m_t$ above $\sim 1.4$ GeV.

The corrected values of $R^{33}_t$, $R^{88}_t$, their difference,
and the resulting values of $m_s$,
are given, as a function of $m_t$, in Table 1.  
$R^{33}_t$ is obtained by combining information from {\rxn} with
that from $\tau$ decay.  In the latter case, 
we have used the recent ALEPH tabulation of the isovector
vector spectral function\cite{ALEPH97},
re-fitting the portion of the 
unfolded distribution from threshold up to $s=(1.6\ {\rm GeV})^2$
relevant to our analysis.  The results turn out to be
almost identical to those obtained by direct numerical
integration of the unfolded ALEPH distribution.
Although the errors associated with the extraction of the isovector
spectral function from $\tau$ decay data are a factor of $\sim 2$
smaller than for {\rxn}, one must bear in mind that
there remain at present unknown
${\cal O}(\alpha_{EM})$ corrections to the relation between
the spectral function extracted in $\tau$ decay
and that appearing in {\rxn}\cite{marciano92}.  
Marciano\cite{marciano92}
has assigned an additional uncertainty of $\sim 3\%$
to account for these corrections.  We have
added this error in quadrature with that quoted by ALEPH.
For the estimate based on {\rxn} data, we employ Narison's
evaluation of the uncorrected hadronic integral for $R^{33}_t$,
and PDG96\cite{PDG96} values for the $\omega$ and $\phi$ EM widths
to obtain the uncorrected version of $R^{88}_t$.
Quoted values for $m_s$ in Table 1 are in the
$\overline{MS}$ scheme, at scale $\mu =1\ {\rm GeV}$, 
using the 4-loop running, and correspond to central values
for all input.
The errors quoted for $R^{88}_t$ reflect both those from the
PDG96 partial widths and those from
the uncertainties in the theoretical analysis of the
mixed-isospin correlator sum rule.  As expected, 
the isospin-breaking corrections have
a very significant impact on $R^{33}_t-R^{88}_t$.
The decrease in the value of the difference
also magnifies the effect of the experimental errors. 
The errors quoted for $m_s$ correspond to
those on the hadronic integrals only
(the first to that on $R^{33}_t$,
the second to that on $R^{88}_t$).  
Errors associated 
with uncertainties in the remaining 
inputs are as follows: (1) for $c_K$:
$\sim \pm 0.4$ MeV; (2) for the light-flavor four-quark
condensate: $\sim \pm 4$ MeV; (3) for $<\bar{s}s>/<\bar{u}u>$: 
$< \pm 17$ MeV; (4) for $f_3$: $<\pm 14$ MeV; (5) for $\Lambda_3$:
$<\pm 3$ MeV.  

We see from Table 1 that
the corrections to $R^{88}_t$
significantly lower
the hadronic side of the sum rule, and hence $m_s$.  
Taking an average of the three
determinations, we obtain
\begin{equation}
\bar{m}_s(1\ {\rm GeV})=113\, {^{+35}_{-54}}\, {^{+32}_{-48}}\pm 23\ {\rm MeV}
\label{final}\end{equation}
the first two sets of (asymmetric) errors being associated with
the experimental input and the last set with
uncertainties in the input on the OPE side.  
The result, while
significantly below that of conventional sum rule analyses,
is compatible with that of Colangelo et al.\cite{colangelo}, and
within errors,
also the low values obtained in some recent lattice 
analyses\cite{bg96,mlatt}.  To understand that the errors associated with
the experimental input are still very significant, note that, if we 
use only the $\tau$ decay data to determine the isovector contributions,
the central values of $m_s$ are changed to $148.5$, $131.5$ and $134.5$
MeV, for $m_t=1.4$, $1.5$ and $1.6$ GeV, respectively.  We should also
point out that the existing extraction of $m_s$ based on flavor-breaking
in hadronic $\tau$ decay\cite{davier} is incorrect as a result of an error
in the input coefficients of the perturbative series given in 
Ref.~\cite{chetyrkin} 
for the mass-dependent D=2 terms entering the strangeness-changing
current contribution to these decays\cite{krmtau}.

\section{Summary}
We have shown that isospin-breaking corrections significantly alter
the value of $\rho^{33}-\rho^{88}$ extracted from existing
{\rxn} data, and that the resulting changes to the Narison
sum rule for $m_s$ produce a central value 
$\bar{m}_s\sim 110$ MeV (at a scale $1$ GeV) ($\sim 140$ MeV if
one uses only $\tau$ decay data for the isovector input to the sum rule).  
The errors on this result, which are not insignificant,
are currently dominated
by the errors on the 
experimental input.
Improved experimental data 
would, of course, significantly reduce the errors on the
hadronic side of the sum rule.  An improved
treatment of both the hadronic and OPE sides of the mixed-isospin
vector current sum rule is also highly desirable, both as a check
of the stability of the results for, and a means of
potentially reducing the errors on the determination 
of, the isospin-breaking vector meson decay
constants.

\acknowledgements
The author would like to thank S. Narison for bringing
the contents of Ref.~\cite{narison95} to his attention, and to acknowledge
both the ongoing support of the Natural Sciences and
Engineering Research Council of Canada, and the hospitality of the
Special Research Centre for the Subatomic Structure of Matter at the
University of Adelaide, where this work was performed.
Useful discussions with Andreas H\"ocker on the ALEPH spectral
function analysis are also gratefully acknowledged.

\begin{table}
\caption{Hadronic input and extracted values of $m_s$}\label{Table1}
\begin{tabular}{ccccc}
$m_t$ (GeV)&$R^{33}_t$&$R^{88}_t$&$R^{33}_t-R^{88}_t$&$m_s$ (MeV)\\
\tableline
1.4&1.890$\pm$ 0.098&1.706$\pm$ 0.057&0.184$\pm$ 0.113&
123.1${^{+37}_{-54}}{^{+23}_{-28}}$\\
1.5&1.842$\pm$ 0.063&1.737$\pm$ 0.056&0.105$\pm$ 0.084&104.4
${^{+36}_{-58}}{^{+33}_{-49}}$\\
1.6&1.800$\pm$ 0.048&1.713$\pm$ 0.056&0.087$\pm$ 0.074&110.4
${^{+33}_{-49}}{^{+39}_{-64}}$\\
\end{tabular}
\end{table}

\end{document}